\documentstyle[a4wide,epsf,11pt,titlepage]{article}

\newcounter{nref}
\newcommand{\bbib}{%
  \renewcommand{\refname}{\large\bf References}%
  \begin{thebibliography}{99}%
  \setcounter{enumiv}{\thenref}}
\newcommand{\ebib}{%
  \end{thebibliography}%
  \setcounter{nref}{\arabic{enumiv}}}
\newcommand{\head}[3]{%
  \setcounter{nref}{0}%
  \thispagestyle{empty}%
  \section*{\LARGE\bf #1}%
  \stepcounter{section}%
  \addcontentsline{toc}{section}{#1}%
  \large\itshape%
  #2\\\vspace{0.1pt}\\%
  #3%
  \normalsize\upshape%
  \bigskip}

\def\yrm1{yr${}^{-1}$}
\def\sun{\odot}
\def\msun{$M_{\sun}$~}

\def\msunend{$M_{\sun}$}

\def\nco{${^{14}N}(e^-,\gamma){^{14}C}(\alpha,\gamma){^{18}O}$~}
\def\3a{3$\alpha$~}
\def\n14{${^{14}}$N}
\def\c14{${^{14}}$C}
\def\o18{${^{18}}$O}

\begin{document}
 
 
\head{The role of NCO reaction chain on the He ignition in degenerate
stellar structures}
{L.\ Piersanti $^{1,2}$, S.\ Cassisi $^{2}$ {\rm \&} A.\ Tornamb\'e $^{2,3}$}
      {$^1$ Dipartimento di Fisica dell'Universit\`a degli Studi di Napoli 
``Federico II'', Mostra d'Oltremare, pad. 20, 80125, Napoli, Italy,
luciano@astrte.te.astro.it\\
      $^2$ Osservatorio Astronomico di Collurania, Via M. Maggini, 64100, 
Teramo, Italy,\\ cassisi@astrte.te.astro.it\\
      $^3$ Dipartimento di Fisica dell'Universit\`a degli Studi de L'Aquila, 
via Vetoio, 67100, L'Aquila, Italy, tornambe@astrte.te.astro.it}

%
%

Some 25 years ago \cite{Mitalas.1} it has been suggested that the \nco
reaction (NCO reaction chain) plays a pivotal role in the onset of the
Helium flash in the core of a low mass Red Giant Branch (RGB) star 
(see \cite{Kaminisi.1} and \cite{Kaminisi.2}). In fact in the He core 
the Fermi energy is large enough to approach the energy for the
e-capture on \n14 well before the \3a reactions produce
energy at a significant level so that a not negligible amount of \c14 is
produced in the most internal regions. In the physical conditions typical of
a star approaching the tip of the RGB $\alpha$ captures on \c14 dominate 
over \3a reactions. In 1980 \cite{Spulak.1} argued that the
NCO reaction does not affect the He ignition in a star at the tip of RGB 
because the physical conditions in the inner region, particularly the density, 
do not allow the NCO reaction to be dominant over the \3a reaction.
Successively, in 1986, Hashimoto, Nomoto, Arai \& Kaminisi
\cite{Hashimoto.1} recomputed the NCO cross sections and found 
that this reaction ``{\ldots} dominates over \3a reaction to heat up the
central region'' of a low mass star approaching the tip of the RGB
\cite{Hashimoto.2}. 

Woosley \& Weaver \cite{Woosley.1} accounted for this reactions chain to
compute the pre-supernova models of cooled down CO WDs accreting He rich
matter. They find that the inclusion of NCO reaction
does not prevent the occurrence of an He detonation. Recently, 
Piersanti, Cassisi, Iben \& Tornamb\'e (\cite{Piersanti.1}) 
have included the NCO chain in the evolution of a low mass CO WD accreting H
and He rich matter at a rate suitable to obtain an He detonation.
Their investigation shows 
that the differences determined by the inclusion of the NCO chain 
are negligible, the final
outcome remaining an explosion. They point out that their result is due to the
fact that they use an evolutionary model, obtained by evolving an
intermediate mass star with moderate mass loss from the Main Sequence phase
till the cooling sequence. In such a model the \n14 abundance at the
base of the He shell, where the He flash takes place, is zero.

In any case it is important to evaluate if the \nco reaction plays
some role in stellar evolution. In fact if this chain
does trigger the onset of the central He flash in low mass star then it 
stops the growth in mass of the He core, modifying the luminosity level of
the RGB tip and of the Horizontal Branch. In addition, if the NCO chain 
plays a role in heating up the He shell in low mass CO WDs 
accreting hydrogen or helium rich matter, then He burning could occur steadily 
allowing the CO core to grow in mass until the Chandrasekhar mass limit.

In order to better understand if this reaction has some role in the
evolution of RG stars we have analyzed three sets of models at different
metallicity (namely Z=0.0001, 0.001 and 0.02): for each sets
we consider three different masses (namely M=0.6, 0.7 and 0.8 \msun). We
follow the evolution of these models from the Pre-Main Sequence phase until
the onset of the He flash at the tip of the RGB. Being the e-capture strongly
dependent on the density and being the maximum density located at the center, 
the physical conditions suitable for the NCO
reaction are attained first at the center of the He core.

We find that for the higher metallicity cases (Z=0.02 and 0.001) the
\3a reaction occurs well before the central density exceeds the
critical value at which the NCO reaction becomes active (${\overline\rho}=10^6$ 
g cm$^{-3}$). Therefore in this case they do not play any role
at all. For models with Z=0.0001 the central density becomes greater 
than $\overline\rho$ before the \3a ignites, therefore  all the central \n14 is
converted into \c14 and the \c14 is converted into \o18, delivering a small
amount of energy that heats up the center. Despite this, the
successive evolution is similar to models which do not 
include the NCO reaction, in the sense that the \3a reaction is ignited 
in the same physical conditions and the final He core mass is the same.

As a whole, we conclude that the NCO chain does not affect at all the
evolution of a low mass star climbing the RGB. In fact for high metallicity
the central density is too low to allow the onset of the electron capture
on \n14; on the contrary, for lower metallicity, the central density
exceeds the critical value for the onset of the NCO reaction but in this
case the \n14 abundance is very low and the produced energy can not heat
up efficiently the structure. 

Our result is in good agreement with \cite{Spulak.1} but it is quite 
different with respect to those obtained by \cite{Hashimoto.2}. This 
occurrence is due to the
fact that Hashimoto and co-workers simulate the behavior of the He core of
a low mass star accreting He rich matter on a cooled down He WD. In this way
they ignore the presence of the overlying H-shell that keeps hot the He core. 
As a consequence, for a fixed metallicity, their models attain a 
too high central density with respect to realistic models.

For what concerns mass accreting CO WDs, it is interesting to note preliminary 
that an evolutionary model presents a CO core surrounded by a He shell and
eventually by an H shell. We have computed three different models from the
central He-burning phase until the cooling sequence, as indicated
in Table \ref{piersanti.tab1}. In the cooled models the \n14
abundance at the physical base of the He shell overlying the CO core is very
small ($< 10^{-10}$ by mass). This is 
due to the fact that during the He-shell burning \n14 is 
converted into \o18 via direct $\alpha$-capture (NO reaction). 
\begin{table}
\begin{center}
\begin{tabular}{c|c|c|c} \hline\hline
{\bf Z} & {\bf Y} & {\bf $M_{tot}$ (\msunend)} & $M_{He-core}$ (\msunend) \\ \hline
0.0001  & 0.244 & 0.520 & 0.498 \\
0.001   & 0.238 & 0.515 & 0.510 \\
0.02    & 0.980 & 0.800 & -     \\ \hline\hline
\end{tabular}
\end{center}
\caption{The main characteristics of the models we have computed from the
central He burning phase until the cooling sequence. Note that the last
model refers to a pure He star.}
\label{piersanti.tab1}
\end{table}

\begin{figure}[ht]
\centerline{\epsfxsize=0.5\textwidth\epsffile{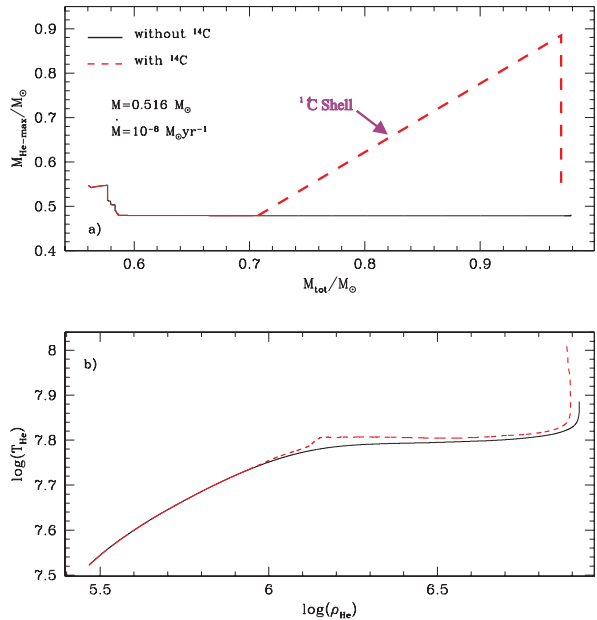}}
\caption{Selected evolutionary properties of a CO WD of 0.516 \msun accreting 
He-rich matter at $\dot{M}=10^{-8}$ \msun\yrm1 with and without NCO reactions. 
{\it Panel a)}: the temporal evolution of the mass coordinate where the energy
production via He-burning is a maximum.
{\it Panel b)}: the evolution in the $\rho - T$ plane of the base of the
He-shell.}
\label{piersanti.fig1}
\end{figure}

The accretion of H or He rich matter causes the growth in mass of the He layer
surrounding the CO core; the He-layers, accreted directly or by H-burning 
by-products, are rich of \n14 so that one can expect that in this case
the NCO reaction could play some role at the ignition of the He burning. 
To verify this scenario we focus the attention on the model of a low mass 
CO WD accreting 
He rich matter at a low rate, suitable for a violent dynamical He ignition 
(see \cite{Piersanti.1}). As it can be seen in Figure \ref{piersanti.fig1}
(panel {\sl a)}) as soon as the physical conditions suitable for NCO burning
are attained, \n14 is completely converted into \c14. The latter element 
undergoes $\alpha$-captures to produce
\o18, heating up locally the structure (see panel {\sl b)}). While the time
elapses, the He layer continues to grow in mass and the \nco 
occur in more and more external zones. Therefore, it is possible to identify the
existence of a \c14 shell which moves outwards. This situation goes on until 
the \3a reactions ignite in high degenerate
physical condition causing the disruption of the accreting structure as
Type Ia supernova. We stress once again that the final outcome remains an
explosion triggered by \3a reactions, the only difference with respect to
the model without NCO chain being a very small reduction in mass 
of the He layers at the runaway.

We conclude that the \nco chain does not play any role at
all in the ignition of the He flash in degenerate and semi-degenerate
physical conditions due to the fact that the ignition density for the
e-capture on \n14 is very high. If the threshold density for the e-capture 
were lower than current value, then the NCO reaction would become of some 
importance. We believe that an up-date of the cross section for 
the e-capture on \n14 is strongly required.

%
%
%
%

\end{document}